\documentclass{ieeeaccess}

\usepackage{cite}
\usepackage{amsmath,amssymb,amsfonts}
\usepackage{algorithmic}
\usepackage{graphicx}
\usepackage{textcomp}
\usepackage{multirow}
\usepackage{caption}
\usepackage{tabularx}
\usepackage{array}
\usepackage{epstopdf}

\begin{document}
\history{Date of publication xxxx 00, 0000, date of current version xxxx 00, 0000.}
\doi{10.1109/ACCESS.2017.DOI}

\title{Quantifying Success in Science: An Overview}
\author{\uppercase{Xiaomei Bai}\authorrefmark{1},
\uppercase{Hanxiao Pan}\authorrefmark{2},
\uppercase{Jie Hou}\authorrefmark{2},
\uppercase{Teng Guo}\authorrefmark{2},
\uppercase{Ivan Lee}\authorrefmark{3},
\uppercase{Feng Xia}\authorrefmark{4}}
\address[1]{Computing Center, Anshan Normal University, Anshan 114007, China}
\address[2]{School of Software, Dalian University of Technology, Dalian 116620, China}
\address[3]{STEM, University of South Australia, Adelaide SA 5001, Australia}
\address[4]{School of Engineering, IT and Physical Sciences, Federation University Australia, Ballarat, VIC 3353, Australia}
\tfootnote{This work was partially supported by Liaoning Provincial Key R\&D Guidance Project (2018104021) and Liaoning Provincial Natural Fund Guidance Plan (20180550011).}


\corresp{Corresponding author: Xiaomei Bai (e-mail: xiaomeibai@outlook.com).}

\begin{abstract}
  Quantifying success in science plays a key role in guiding funding allocations, recruitment decisions, and rewards. Recently, a significant amount of progresses have been made towards quantifying success in science. This lack of detailed analysis and summary continues a practical issue. The literature reports the factors influencing scholarly impact and evaluation methods and indices aimed at overcoming this crucial weakness. We focus on categorizing and reviewing the current development on evaluation indices of scholarly impact, including paper impact, scholar impact, and journal impact. Besides, we summarize the issues of existing evaluation methods and indices, investigate the open issues and challenges, and provide possible solutions, including the pattern of collaboration impact, unified evaluation standards, implicit success factor mining, dynamic academic network embedding, and scholarly impact inflation. This paper should help the researchers obtaining a broader understanding of quantifying success in science, and identifying some potential research directions.

  \end{abstract}


\titlepgskip=-15pt

\maketitle
\section{Introduction}
\label{sec:introduction}
Success in science refers to scientists' achievements in their academic careers. Quantifying success in science has developed into a very important part of bibliometrics and scientometrics. An influential publication or scholar always brings much to the followers to carry out their research. Therefore, the ability of bibliography retrieval is very important for researchers, including mining, managing, and examining scholarly big data to identify the successful papers and scholars \cite{xia2017big,liu2018artificial,wang2017shifu,wang2019sustainable,he2020the}. In addition, quantifying scholar impact has special significance in funding allocation and recruitment decisions. Quantifying the impact of paper and journal can help scientists know the frontier of science development. Therefore, quantifying success in science provides useful guidance to the scientific community, such as offering candidates to university, recommending scientists for promotion, and distribution for research funds \cite{Feng2016Scientific,xia2014mvcwalker}.

Quantifying success in science mainly focuses on quantifying the current impact of academic entities, including paper, scholar, journal, scholarly team, and institution \cite{bai2017role,bai2017overview,amjad2020scientific,bai2020measure}. Because the research on the impact of paper, author, and journal is very rich, this paper mainly introduces quantified success in science from these three aspects. Generally, the number of citations is used as an evaluation indicator, which is derived from its easy availability. Lots of factors influence a paper's success, such as paper's visibility \cite{c2-3,liu2018survey} and paper's age \cite{c2-4}. A common method to judge the success of a scholarly paper is to use evaluation indicators, which may take into several important factors. The counting-based and network-based evaluation methods are frequently used to quantify success in science. The counting-based methods are the most direct representation of evaluating, such as citations, author's h-index \cite{c2-62}, and Journal Impact Factor (JIF) \cite{Garfield1972Citation}. Different academic entities form different kinds of academic networks, such as citation network, co-author network, and co-citation network \cite{chen2019analytic}. Currently, the HITS-type and PageRank-type algorithms can mine the complex scholarly relationship based on different scholarly networks and give reasonable evaluation. The features of scholarly networks are also critical to evaluate paper impact. Further, based on these features, many researchers have improved PageRank \cite{Page1998The} or HITS \cite{Kleinberg1999Authoritative} algorithms to make them more suitable to measure the impact of paper.

Same as quantifying the impact of paper, scholar impact is also influenced by many factors. Lots of methods and indices to measure scholar impact are proposed, such as h-index \cite{c2-33}, g-index \cite{c2-34}, and hg-index \cite{Alonso2010hg}. These indices can be unfair for some young researchers because the quality and quantity of a scholar's publications are associated with their academic ages. The methods based on the network can avoid this situation to a certain extent.

Evaluating journal impact is an important part of quantifying success in science.
Many network-based evaluation methods and indices are used to quantify the impact of paper and author, which can also be used to evaluate journal \cite{c2-81,c2-84,c2-85,c2-86}. These methods are based on PageRank, HITS, or consider the structural position of a journal in the journal citation network. In addition, Journal Citation Reports (JCR) is very popular for ranking journals.

Even though the existing research provides a tool to quantify success in science, it still has some limitations. Every indicator to quantify scientific impact has its shortages. In particular, in quantifying scientific success research, one of the most challenging problems stems from the heterogeneous attribute and the dynamic nature of big scholarly data. At present, in most of quantifying scientific success methods, implicit features and implicit relationships have attracted the attention of researchers \cite{bai2018quantifying}.

This paper presents a review of recent developments in quantifying success in science and this review complements relevant work in the past: Wildgaard et al.~\cite{Wildgaard2014A} present a review on author impact evaluation. One limitation of this review is that it does not consider paper and journal impact evaluation research. Bai et al.~\cite{bai2017overview} offer a review of the literature on paper impact evaluation. This overview covers key techniques and paper impact metrics. The limitation of this work is that authors have not consider author and journal impact evaluation. In addition, factors influencing scholarly impact have not analyzed. Therefore, in this paper, the progress of impact evaluation of the paper, author and, journal is described in detail.

\begin{figure*}[htbp]
  \begin{center}
    \includegraphics[width=0.8\textwidth]{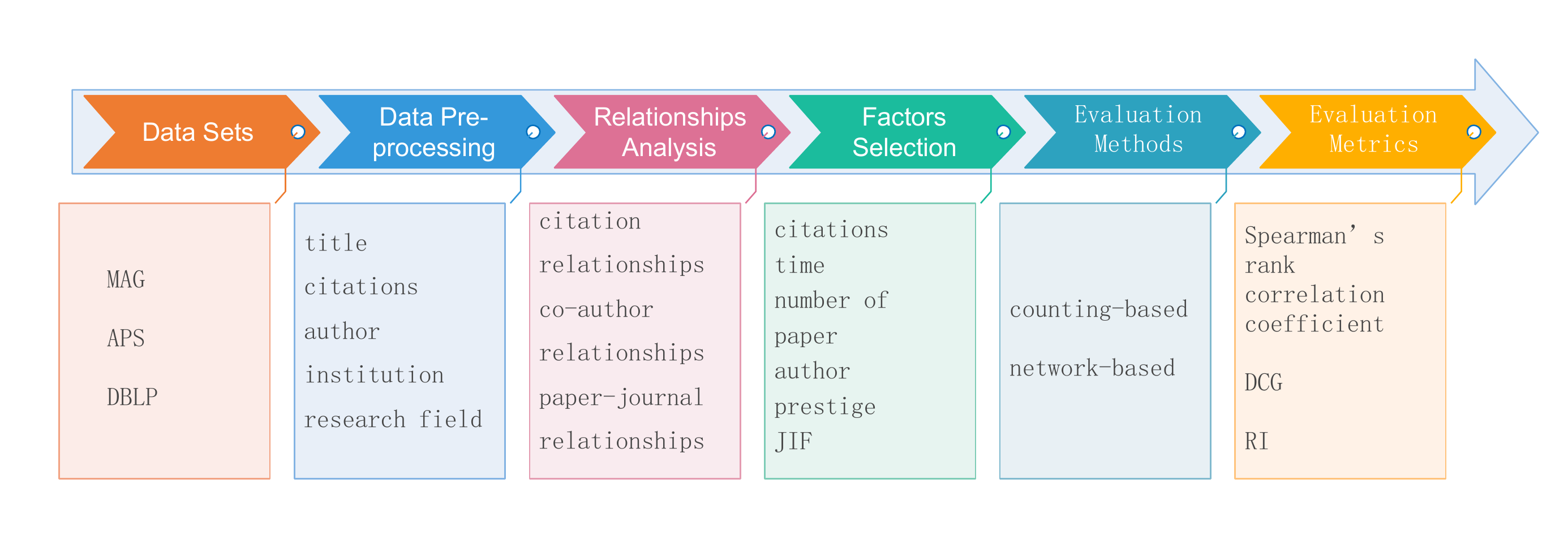}
    \caption{Framework of quantifying success in science.}
    \label{fig1}
  \end{center}
\end{figure*}

FIGURE \ref{fig1} shows the framework of quantifying success in science. Quantifying success in science includes the following parts: data collection, data pre-processing, relationship analysis, evaluation method and evaluation indices. Several public accessible data sets are used to quantify success in science, including American Physical Society (APS)\footnote{http://publish.aps.org}, Digital Bibliography \& Library Project (DBLP)\footnote{https://dblp.uni-trier.de/}, and Microsoft Academic Graph (MAG)\footnote{http://aka.ms/academicgraph}. Data pre-processing in quantitative scientific success studies is very important because it relates to the accuracy. The homogenous and heterogeneous scholarly networks are used to research the scholarly relationships such as citation relationships, co-author relationships, and paper-journal relationships. Spearman's rank correlation coefficient, Discounted Cumulative Gain, and RI can be used as evaluation metrics for quantifying success in science \cite{Bai2019Author,Bai2016Identifying}. Specially, the heterogeneous scholarly network structures have increased the challenges in scholarly network analysis.

To retrieve the papers of quantifying success in science, based on Google Scholar, we enter search terms such as the success of science, paper impact, scholar impact, journal impact, etc. We first search for the related papers recently published in top journals and top conferences, and then look for their references, and the papers cite these papers to obtain more related papers. Search for papers in a step-by-step manner, then filter and classify from three aspects: paper impact, scholar impact and journal impact, and retain the representative related papers. Based on the above work, we mark the publication years of these papers, read through these papers by year, analyze and summarize the following aspects: the features that influence scholarly impact, evaluation methods and indices. For example, in terms of these features of evaluation paper impact, we classify these features, including reference, references, selected features, statistical feature, network feature, explicit feature, implicit feature, and evaluating paper impact. By analyzing and summarizing these evaluation methods, we identify open issues and challenges, and provide possible solutions.

The rest of this survey is organized as follows. In Section \ref{sec:2.1}, we discuss the evaluation of paper impact. In Section \ref{sec:2.2}, we introduce the evaluation of author impact. The evaluation of journal impact is discussed in section \ref{sec:2.3}. Open issues will be discussed in Section \ref{sec:4}. Finally, we conclude this survey in Section \ref{sec:5}.
\section{Evaluation of Paper Impact}
\label{sec:2.1}
In this section, we will make a detailed introduction to the evaluation methods and indices of paper impact. Besides, we will discuss the evolution of the existent methods and indices, showing their advantages and shortcomings. At first, we begin with the evaluation of paper impact, because many assessment methods and indices of scholars and journals are based on the assessment of their papers. Therefore, it is of great significance whether the quality of papers can be quantified accurately. Although the value of a paper is mainly based on its content, the evaluation of its content is easily influenced by subjective factors, and the evaluation efficiency cannot meet the demand of scholarly bid data. This phenomenon drives researchers to give some accurate, efficient automatic evaluation methods. One possible solution is to construct a multi-dimensional metric in which the importance of citation, social relationships of authors, the relationship between the impact of early citers and scholarly paper impact, and citation inflation need to be explored.
\subsection{Factors Influencing the Impact of Papers}
\label{sec:2.1.1}
TABLE \ref{tab:1} shows an example of selected features for evaluating paper impact, including references, selected features, statistical feature, network feature, explicit feature, implicit feature, and evaluating paper impact.
\begin{table*}[!htbp]
  \centering
  \caption{An example of selected features for evaluating paper impact.}
  \begin{tabular}{|p{1.5cm}<{\centering}|p{3cm}<{\centering}|p{1.5cm}<{\centering}|p{1.5cm}<{\centering}|p{1.5cm}<{\centering}|p{2.5cm}<{\centering}|p{2.5cm}<{\centering}|}
    \hline
    References&Selected Features&Statistical Feature& Network Feature& Explicit Feature&Implicit Feature&Evaluating paper impact \\ \hline
      ~\cite{c2-4} &citation rate of a paper, time&yes& no &yes&no&the citation rate of a paper at a given time\\ \hline
      ~\cite{bai2018quantifying}&a relative citation weight &yes& no &no&yes& applying a relative citation weight to the higher-order quantum PageRank algorithm\\ \hline
      ~\cite{Bai2016Identifying} &collaboration times, the time span of collaboration, citing times and the time span of citing&yes& no &yes&no& weakening the relationship of Conflict of Interest (COI) in the citation network\\ \hline
      ~\cite{Lehmann2006Measures} &number of citations& yes & no & yes & no& using the number of citations\\  \hline
      ~\cite{c2-22} &citations, authors, journals/conferences and the publication time information &no& yes &yes&yes& integrating the selected features into PageRank and HITS algorithms\\ \hline
    ~\cite{Wang2013Quantifying} &preferential attachment, aging, fitness&yes& no &no&yes&identifying the three fundamental mechanisms to evaluate long-term impact \\ \hline
    ~\cite{c2-14}&importance of paper&no&yes&yes&no&applying the Google PageRank algorithm to obtain the relative importance of all publications\\ \hline
    ~\cite{Zhang2019Ranking}&citation relevance and author contribution&no&yes&yes&no&using the selected features to weight citation network and  authorship network to evaluate paper impact\\ \hline
     ~\cite{piwowar2013altmetrics}&Altmetrics &yes& no &yes&no& monitoring citations, blogs,
tweets, download statistics and attributions in research articles\\ \hline
     ~\cite{c2-21}&prestige of a paper, prestige of author, time&yes& yes &no&yes& using the citation network, the authorship
network and the publication time of the article for predicting future citations\\ \hline
 ~\cite{wang2019quantifying}&the time-weighted citation count, the citation width, the citation depth&yes& no &no&yes& using entropy weight the three indices\\ \hline
    \end{tabular}
  \label{tab:1}
\end{table*}

The number of citations has been used as a metric to evaluate paper impact for a long time~\cite{Lehmann2006Measures}. Since the number of citations is relatively easy to obtain, it is frequently manipulated such as self-citation, mutual citation, and friend's citation. Although some scholars can cite their papers, because their research subjects can have several stages output and the former results can be the foundation of the latter. But if a self-citation only means to increase the number of citations, it will mislead the scholarly evaluation and bring unfair factors to the evaluation system. For inappropriate citations, previous researchers proposed corresponding methods to waken the influence of self-citation by relying on the higher-order citation network~\cite{bai2018quantifying}.

Previous research shows that the impact of paper will decay over time, which confirms that the age of a paper is a factor influencing its impact. Generally, an old paper has more citations than a new one, but its work was already covered by new papers so that it could get fewer citations in the future. Parolo et al.~\cite{c2-4} showed that the decay of the attention paid to a paper is a universal phenomenon, and the decay rate is close to a power law. In some cases, papers can be forgotten more quickly so the attention decay is faster, which fits an exponential curve. The time factor, the prestige of a paper, and the prestige of the author were used to evaluate scholarly paper impact~\cite{c2-21}. Based on the three factors, they evaluated scholarly paper impact by predicting the number of citations of scholarly papers in the future.
Wang et al.~\cite{Wang2013Quantifying} considered the aging factor to evaluate paper impact because it can capture the fact that new ideas are integrated in subsequent work. Wang et al.~\cite{wang2019quantifying} first developed the three indices: the time-weighted citation count, the citation width, the citation depth. They then leveraged entropy to weight these indices to evaluate paper impact.
Chan et al.~\cite{c2-5} discussed that the impact of authors and affiliations can influence on the impact of their papers. In their research, they argued that the reputation of authors and the impact of their affiliations had the power to boost paper impact in the early stages of publication, but this influence could decay fast and in the following stages. Chen et al.~\cite{c2-14} found the scientific gems using Google's PageRank algorithm in the citation network. Zhang et al.~\cite{Zhang2019Ranking} evaluated the impact of authors and papers based on the heterogenous author-citation academic networks.

In addition to the factors mentioned above, some other factors were also used to evaluate paper impact, such as individual, institutional and international collaboration, reference impact, reference totals, keyword totals, and abstract readability~\cite{c2-7}. Preferential attachment, fitness, and aging factors were used to quantify the long-term scientific impact, and the three factors can drive the citation history of scholarly paper~\cite{Wang2013Quantifying}. In this research, the preferential attachment captures the fact that highly cited papers are more likely to be cited again than less-cited papers. Fitness captures the inherent differences between scholarly papers. The aging has been introduced before. It can be traced back to the journal impact factor that was once used as a criterion for assessing the impact of a paper \cite{Seglen1997Why}. Altmetrics evaluated scholarly impact based on the activities in the social media platforms, such as citations, blogs,
tweets, download statistics, and attributions in research articles~\cite{piwowar2013altmetrics}. Altmetrics scores were used to complement the evaluation of scholarly paper with new insights~\cite{costas2015altmetrics}.
Since we have already known most factors that influence the impact of paper, the evaluation methods and the corresponding indices can be designed.
\subsection{Counting-Based Evaluation Methods and Indices}
\label{sec:2.1.2}
TABLE \ref{tab:2} shows the comparison of counting-based evaluation methods and indices from the following aspects: method and reference, selected factors, importance of each citation, advantage and disadvantage.
\begin{table*}[!htbp]
  \centering
  \caption{An example of counting-based method comparison for evaluating paper impact.}
  \begin{tabular}{|p{3cm}<{\centering}|p{2cm}<{\centering}|p{2cm}<{\centering}|p{2cm}<{\centering}|p{4cm}<{\centering}|}
    \hline
    Method and Reference&Selected Factors& Importance of Each Citation &Advantage&Disadvantage\\ \hline
    citations~\cite{Price1965Networks}& citations&equal&Easy to obtained.& Easy to be manipulated. Strong despondence on paper's age.\\  \hline
    impact factor~\cite{Seglen1997Why}& number of paper, citations of paper, time & equal & Easy to calculate. & Easy to be manipulated. Hard to unify impact factors across different disciplines.\\  \hline
    a SVR model~\cite{c2-9}& occurrence times, located section, time interval, self-cited& unequal & Distinguish the importance of citation. & Hard to calculate. \\  \hline
    a supervised machine learning model~\cite{Zhu2015Measuring}& citation location, semantic similarities, cited frequency, number of citations& unequal & It can distinguish the importance of citation. &Hard to calculate. \\  \hline
    paper's normalized distribution of citation and JIF~\cite{c2-10}& distribution of citation, JIF& equal & It is feasible on a scale typical of a national evaluation exercise. & Easy to be manipulated. \\  \hline

  \end{tabular}
  \label{tab:2}
\end{table*}

Garfield et al.~\cite{c2-60} first proposed using the number of citations to assess the impact of scholarly papers.
 Citations are the simplest and most direct counting-based index of paper impact. However, citations as an evaluation metric have some drawbacks. For example, it relies heavily on the time of publication of the paper. The longer the time is, the more the citations are. Considering this drawback, previous research used the journal impact factor to quantify the impact of paper~\cite{Seglen1997Why}. The reason is that to a certain extent, journal impact can characterize paper impact. However, Seglen et al.~\cite{Seglen1997Why} summarized problems associated with the use of journal impact factors, and they found that the journal impact factor is not representative of individual paper.
It has been recognized that not all citations are equal importance and hence the importance of citation needs be distinguished~\cite{Zhu2015Measuring}.

To distinguish the importance of citation, previous researchers have made many attempts. Wan et al.~\cite{c2-9} divided the importance of citation into 5 levels, which was called citation strength. In their research, the importance of citation was determined by the following features: occurrence times, located section, time interval, the average length of citing sentences, average density of citation occurrences, and self-cited. Then a SVR model was used to calculate every citation's importance level with giving some artificially labeled data. The impact of a paper is calculated by summing up all the citation strengthes. Their experimental results showed that ranking papers using citation strength fitted the ground truth better. Zhu et al.~\cite{Zhu2015Measuring} distinguished the importance of citation by identifying a set of four features that are useful to determine the impact of a scholarly paper, including citation location in paper, semantic similarities between titles of cited paper and the content of citing paper, cited frequency, number of citations in a literature.

Anfossi et al.~\cite{c2-10} argued that it was more reasonable to rank papers by combining the information of several indicators than using only one. In their paper, an evaluation tool was proposed, which used paper's normalized distribution of citation and JIF and located a paper in the (citation, JIF) space intuitively as a scatter plot. Then this space was divided into regions by drawing thresholds as weighted linear combinations of the paper's citation and JIF, shown in Function $(1)$,
\begin{equation}
\begin{split}
    f_n(CIT,JIF) = Const_n + a_{1n} \cdot CIT + a_{2n} \cdot JIF + \\
    a_{3n} \cdot CIT \cdot JIF + a_{4n} \cdot CIT^2 + a_{5n} \cdot JIF^2 + \cdots
\end{split}
\end{equation}
where $const_n$ is a constant that controls the segmentation of the region, and CIT indicates paper's citation. The different calibrations of the segmentation result in different classifications of articles. Before Anfossi's work, Ancaiani et al.~\cite{c2-11} performed an analysis of a large amount of research outcomes submitted by Italian universities and other research bodies.

Nowadays more and more research results or papers are spreading on social media, which is helpful to promote a scholar's impact. The times of downloading, sharing, or commenting of papers on the online social networks have already been a group of metrics to evaluate the research outputs, which are known as Altmetrics \cite{piwowar2013altmetrics}. The social network-based Altmetrics are used more and more widely as a new emerging evaluation metrics of paper. Xia et al.~\cite{c2-13} performed an analysis on how the Twitter and Facebook users impact the paper's influence published on \textit{Nature}. They found that the users of Twitter are easier to spread the impact of papers published on \textit{Nature}. Although Altmetrics are able to complement and improve the assessment of paper impact, Altmetrics are not authoritative as an evaluation indicator. Mainly because Altmetrics are easily manipulated as citations. The method of quantifying academic impact based on Altmetrics needs further exploration.
\subsection{Network-Based Evaluation Methods and Indices}
\label{sec:2.1.3}
TABLE \ref{tab:3} shows the comparison on network-based evaluation methods and indices from the following aspects:
method and reference, selected factors, scholarly network, algorithms, advantage and disadvantage.
\begin{table*}[!htbp]
  \centering
  \caption{An example of network-based method comparison for evaluating paper impact.}
  \begin{tabular}{|p{2.5cm}<{\centering}|p{2cm}<{\centering}|p{2cm}<{\centering}|p{2cm}<{\centering}|p{1.5cm}<{\centering}|p{2.5cm}<{\centering}|p{2.5cm}<{\centering}|}
    \hline
    Method and Reference&Scholarly Network& Homogeneous Network& Heterogeneous Network&Algorithms &Advantage&Disadvantage\\ \hline
    PageRank~\cite{c2-14}&citation network&yes &no &PageRank& It begins to use a structured approach to quantify paper impact. & It does not consider attenuation of paper impact over time. \\  \hline
     CiteRank~\cite{c2-16}&citation network&yes&no&PageRank& It promotes the impact of recent publications. & It does not consider the impact of author and journal. \\  \hline
     nonlinearity PageRank~\cite{c2-17}&citation network&yes&no&PageRank& This method can control the paper's score accumulation. &It does not consider the impact of author and journal. \\  \hline
     PageRank-type method~\cite{c2-23}&co-author network, citation network, author-paper network, paper-text feature network,author-text feature network&yes&yes&PageRank& This method can control the paper's score accumulation. &It does not consider the impact of author and journal. \\  \hline
     HITS-type method~\cite{c2-19}&citation network,co-author network&yes &yes&HITS&This method can evaluate paper and its author at the same time.&It does not consider the impact of journal. \\  \hline
     Tri-Rank~\cite{c2-20}&citation network,co-author network, venue citation network&yes &yes&HITS&This method can rank authors, papers and venues simultaneously in heterogeneous networks.&It does not consider attenuation of paper impact over time.\\  \hline
      Future-Rank~\cite{c2-21}&citation network, paper-author network&yes &yes&PageRank, HITS& FutureRank can combine information about citations, authors and publication time to rank papers.&It does not consider the impact of journal.\\  \hline
      CAJTRank~\cite{c2-22}&citation network, paper-author network, paper-journal&yes &yes&PageRank, HITS& It can combine information about citations, authors, journal and publication time to rank papers.& Citation weights are equal.\\  \hline
      COI-Rank~\cite{Bai2016Identifying}&citation network, paper-author network, paper-journal&yes &yes&PageRank, HITS& It can distinguish the importance of citation in heterogeneous scholarly network.& COI relationship contains many factors, and it is not easy to mine.\\  \hline
      higher-order weighted quantum PageRank~\cite{bai2018quantifying}&citation network&yes &no&Quantum PageRank& It can reveal the actual impact of papers, including necessary self-citations.& Time is costly.\\  \hline
  \end{tabular}
  \label{tab:3}
\end{table*}

\begin{figure*}[htbp]
  \begin{center}
    \includegraphics[width=0.7\textwidth]{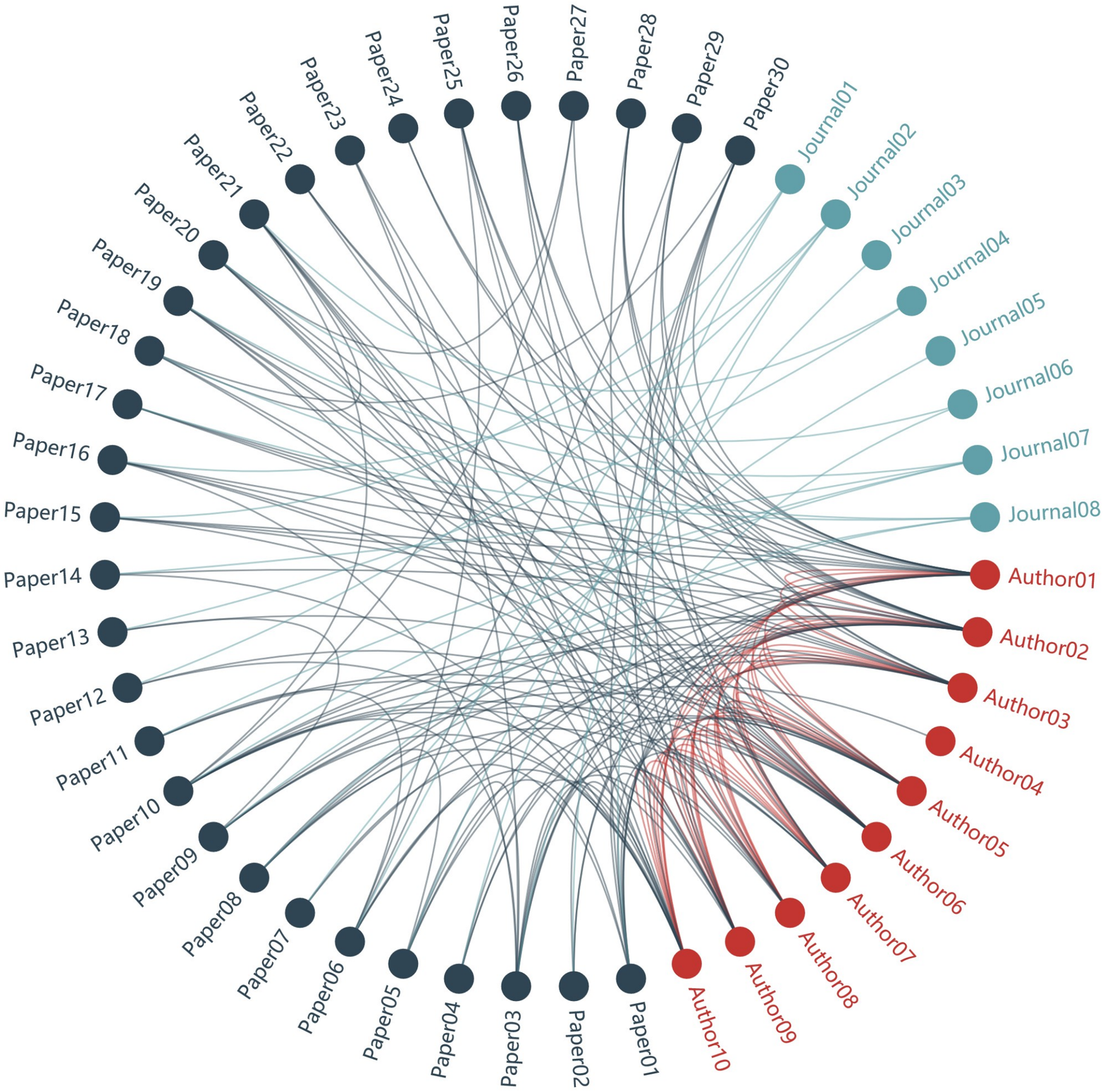}
    \caption{Several typical scholarly networks for paper impact evaluation.}
    \label{fig2}
  \end{center}
\end{figure*}

A classical network-based evaluation method is PageRank algorithm~\cite{Page1998The}. Another famous algorithm for evaluating the importance of nodes in heterogeneous networks is HITS. The two methods have been used to quantify the impact of papers. PageRank algorithm is used in a homogeneous scholarly network, and HITS is used a heterogeneous scholarly network. FIGURE \ref{fig2} shows several typical scholarly networks for paper impact evaluation, such as citation network, co-author network, paper-author network, paper-journal network. The four scholarly networks are generated from randomly selected 10 authors for the computer science area in the MAG dataset. The different color nodes indicate different types of academic entities and the lines between them indicate their scholarly relationships.

Chen et al.~\cite{c2-14} applied the Google PageRank algorithm on all publications in the Physical Review family of journals from 1893 to 2003 to find out some exceptional papers. PageRank can find the linear relation among papers in the citation network. Recently, London et al.~\cite{c2-15} proposed a local form of PageRank to evaluate the impact of paper only on a small set of nodes extracted from the whole citation network. A paper that has more citations or has been cited by an important paper will be set a higher score through the algorithm. But the classical PageRank algorithm is non-time-sensitive. This leads to an unreasonable result that an out-of-date paper may still get a high impact because of its citations accumulating long before, but its true value has already been replaced by many new publications. To overcome this problem, Walker et al.~\cite{c2-16} introduced CiteRank, to weight with time-based on PageRank to promote recent publications. The function of this method is as follows:
\begin{equation}
\begin{split}
    T = I \cdot \rho + (1 - \alpha)W \cdot \rho + (1 - \alpha)^2 W^2 \cdot \rho + \cdots
\end{split}
\end{equation}
$T$ is a matrix of the final scores of all papers. $W$ is the transferring probability matrix where $W_{ij} = 1 / k_j^{out}$ if $j$ cites $i$ and 0 otherwise, where $k_j^{out}$ is the out degree of the $j$th paper. $\rho_i$ is the initial probability of selecting the $i$th paper in the citation network, there given as $\rho_i = e^{-age_i / \tau_{dir}}$, where $age_i$ indicates years the $i$th paper's after published.

Many efforts have been paid for updating the PageRank to make it fit the characteristics of the academic network. Yao et al.~\cite{c2-17} introduced nonlinearity to the PageRank algorithm by aggregating the score from downstream neighboring nodes in a nonlinearity way. The iteration function changes into the following form correspondingly:
\begin{equation}
\begin{split}
    s_i(t) = \alpha + (1 - \alpha) \bigg[  \sum_{j = 1}^n \frac{1}{N} \delta_{k_j^{out},0} s(t - 1) + \\
    \sqrt[\theta + 1]{\sum_{j = 1}^n A_{ij} (1 - \delta_{k_j^{out},0})(\frac{s_j(t - 1)}{k_j^{out}})^{\theta + 1}} \bigg]
\end{split}
\end{equation}
By tuning the value of $\theta$, this method can control the paper's score accumulation and make it more sensitive to the citer's impact. This nonlinear method considers that the value of a citation from high impact paper is more important than the one from low-level paper.

Wang et al.~\cite{c2-23} proposed a PageRank-type method that used several scholarly networks to rank papers, including a time-aware co-author network ($M^{AA}$), a time-aware paper citation network ($M^{PP}$), an author-paper network ($M^{AP}$) indicating the paper's authorship, a paper-text feature network ($M^{PT}$) indicating the paper's textual features and an author-text feature network ($M^{AT}$). The iteration equation is:
\begin{equation}
\begin{split}
    R^{t + 1} = MR^t,
\end{split}
\end{equation}
where $R = [A\_P^T, A\_A^T, A\_F^T]^T$, and  M =
\begin{displaymath}
  \left(
        \setlength{\arraycolsep}{0pt}
        \begin{array}{ccc}
            \alpha_pM^{PP}\Lambda_I & \beta_p(1-\alpha_p)M^{PA} & (1-\beta_p)(1-\alpha_p)M^{PT}\\
            \beta_{\alpha}M^{AP} & \alpha_{\alpha}M^{AA}\Lambda_I & (1-\beta_{\alpha})(1-\alpha_{\alpha})M^{AT}\\
            (1-\alpha_f)\Lambda_EM^{PT} & \alpha_f\Lambda_EM^{TA} & \Lambda_0 \\
        \end{array}
    \right).
\end{displaymath}
$\Lambda_I$ and $\Lambda_E$ are both diagonal matrixes with the diagonal elements $\Lambda_{ii} = 1$ and $\Lambda_{ii} = E_i$, respectively. $\Lambda_0$ is a zero matrix. Vectors $A\_P^T, A\_A^T$ and $A\_F^T$ are authority of paper, author and text features respectively.

Jiang et al.~\cite{c2-18} took this dynamic evolution of citation network into account and put forward a method with the same idea of PageRank. The method integrates three factors in scientific development, including knowledge accumulation by individual papers, knowledge diffusion through citation behavior, and knowledge decay with time elapse. Then it uses a random walk process on the citation network to describe these three factors. The dynamically evolving process is simulated by dividing all papers according to their publishing time and adding into the citation network partially with the time sequence.

Another type of method is based on HITS~\cite{Kleinberg1999Authoritative}. Zhou et al.~\cite{c2-19} performed the HITS algorithm on paper's citation network and co-author network, which were connected by authorship. In both citation network and co-author network, nodes' scores were first calculated by PageRank, and then a HITS was performed on the bipartite graph to get the final scores of papers and authors. So this method can evaluate the impact of authors and their papers at the same time. The iteration function is as follows:
\begin{equation}
\begin{split}
    \textbf{a}_{t+1} = (1 - \lambda)(\widetilde{A}^T)^m \textbf{a}^t + \lambda DA^T(AD^TDA^T)^k \textbf{d}^t \\
    \textbf{d}_{t+1} = (1 - \lambda)(\widetilde{D}^T)^n \textbf{a}^t + \lambda AD^T(DA^TAD^T)^k \textbf{a}^t,
\end{split}
\end{equation}
where matrix $A$ and $D$ are the transferring probability matrix of co-author network and citation network correspondingly. And $\widetilde{A}$ is the iteration matrix in the PageRank process on co-author network, which is given by $\widetilde{A} = (1 - \alpha)A + \frac{\alpha}{n_A}\mathbb{I}$, where $\mathbb{I}$ is a matrix with all elements equaling 1. $\widetilde{D}$ is the same meaning. Vector $\textbf{a}$ storages the scores of all authors and vector $\textbf{d}$ storages the scores of all papers. A similar method is the Tri-Rank algorithm proposed by reference~\cite{c2-20} in 2014, which took the paper's publication information into account and performed a HITS-type method on three linked networks, adding a venue citation network on the two networks used before.

In addition, some methods that combine PageRank and HITS to evaluate the impact of papers. A typical one is FutureRank, proposed by reference~\cite{c2-21}. Different from other methods, FutureRank ranks the impact of papers and authors by predicting their future PageRank scores. PageRank algorithm is first used to rank papers via the citation network, and then the HITS algorithm is used to calculate the authority score of papers and hub score of authors based on the hybrid network. After calculating the PageRank score of papers, the authority score of papers, and the hub score of authors, the final result of the evaluation is finally obtained by weighting to these scores, seeing function (6).
\begin{equation}
\begin{split}
    S(P_i) = \alpha * PageRank(P_i) \\
           + \beta * Authority(P_i) \\
           + \gamma * Hub(P_i) \\
           + (1 - \alpha - \beta - \gamma) * 1/n,
\end{split}
\end{equation}
where $n$ is the number of nodes in the network. Wang et al.~\cite{c2-22} proposed a similar method that added a journal/conference network to show where the paper was published. The evaluation method's form is the same as FutureRank but it can rank journals/conferences together. Using the HITS algorithm can also evaluate paper and author's quality. Based on their work, Bai et al.\cite{Bai2016Identifying} ranked scholarly papers by investigating the citation relationships to weaken the relationship of Conflict of Interest in the citation network. To a certain extent, this method weakens the impact of self-citation. Besides, Bai et al.\cite{bai2018quantifying} quantified the impact of scholarly papers based on the higher-order weighted citations. In this research, a higher-order weighted quantum PageRank algorithm is developed to reflect the multi-step citation behavior. One advantage of the method is that it can weaken the effect of manipulated citation activities.

\section{Evaluation of Scholar Impact}
\label{sec:2.2}
The evaluation of scholars always relates to their papers. Many methods can evaluate paper together with its authors, such as Co-rank \cite{c2-19}, Tri-Rank \cite{c2-20}, FutureRank \cite{c2-21}, s-index \cite{c2-26}. These network-based methods usually rank several academic entities together because using information provided by a single network is always not enough to give a reasonable evaluation. There are also some counting-based evaluation methods like the famous h-index for quantifying author impact. In this section, we compare different counting-based methods, including method and reference, selected factors, importance of each citation, advantage, and disadvantage. We also compare different network-based methods based on the following several aspects: method and reference, scholarly network, homogeneous network, heterogeneous network, algorithms, advantage, and disadvantage. Besides, we will discuss the evolution of the existent methods and indices and summarize the issues of these methods. One possible solution is to explore the higher-order academic network analysis, author impact inflation, and academic success gene.
\subsection{Factors Influencing the Impact of Scholars}
\label{sec:2.2.1}
The author impact evaluation has undergone a transition from unstructured measure to structured measure \cite{Bai2019Author}. The factors used by researchers to assess author impact ranged from simple statistical factors to structural factors, from explicit factors to implicit factors. Currently, the commonly used factors influencing the impact of scholars can be divided into six categories including paper-related, author-related, venue-related, social-related, reference-related, and temporal-related factors. TABLE \ref{tab:4} shows an example of selected factors for evaluating
author impact.
\begin{table*}[!htbp]
  \centering
  \caption{An example of selected factors for evaluating author impact.}
  \begin{tabular}{|p{3.0cm}<{\centering}|p{2cm}<{\centering}|p{2cm}<{\centering}|p{2cm}<{\centering}|p{2.5cm}<{\centering}|}
    \hline
    Factors&Factor category&Explicit Factor&Implicit Factor&References\\ \hline
    number of citations&paper-related&yes & no& \cite{Bouyssou2016Ranking,marchant2009score} \\  \hline
    the number of publications&paper-related&yes & no& \cite{marchant2009score,c2-39} \\  \hline
    papers scores&paper-related&yes & no& \cite{usmani2017unified} \\  \hline
     share keywords between author and paper &paper-related&yes & no& \cite{zhang2018task} \\  \hline
     PageRank&paper-related&no & yes& \cite{senanayake2015pagerank,Dunaiski2018Author} \\  \hline
     paper authority vector&paper-related&no & yes& \cite{c2-25} \\  \hline
        the number of author&author-related&yes & no& \cite{marchant2009score} \\  \hline
     Maximum Entropy&author-related&yes & no& \cite{c2-39} \\  \hline
      venues scores&venue-related&yes & no& \cite{usmani2017unified} \\  \hline
     journal impact factor&venue-related&yes & no& \cite{c2-39} \\  \hline
    paper's references being cited by the author before, ratio of the paper's references being cited by the author before, paper's references in the author's previous publications&reference-related&yes & no& \cite{zhang2018task,dong2015will} \\  \hline
    times the author attend the paper's venue before, ratio of times the author attend the paper's venue before&reference-related&yes & no& \cite{zhang2018task} \\  \hline
   time&time-related&yes & no& \cite{zhang2017exploring} \\  \hline
  \end{tabular}
  \label{tab:4}
\end{table*}

In the scientific community, scholars can continuously accumulate academic impact but to some extent, the inherent impact of scholars determines their final research results. Since the papers published by scholars can represent the impact of scholars, the paper-related factors are frequently used to measure the impact of scholars. These factors can be selected primarily to consider the quality and quantity of the papers. However, these factors can lead to bias. The academic output of scholars is generally related to their academic age. Scholars with an old academic age may have more output. In this way, simply evaluating scholar impact in terms of output has a big bias for newcomers. Such biases also exist when evaluating scholar impact across research fields. Scientists have made many attempts to eliminate the imbalance between disciplines in evaluating scholar impact. In addition, the allocation of contributions of co-authors of a scholarly paper may also lead to bias in scholar impact evaluation. Shen et al.~\cite{c2-30} developed a credit allocation algorithm to capture the co-authors' contributions.

To a certain extent, author-related factors and venue-related factors can reflect the scholar's impact. Dong et al.~\cite{dong2015will} found that two factors, the impact of scholars and venues, played a key role in improving the h-index of lead authors. Deville et al.~\cite{c2-31} discussed the mobility patterns of scientists at an institutional level and success in science in their careers. They found that the consequence of scholars switching from high-impact institutions to low-impact institutions is a decline in both research quality and output, suggesting that the academic environment has an impact on academic outcomes. Scholars also use online platforms (Google Scholar, Microsoft Academic Search), and social media to enhance their academic impact. Mas-Bleda et al.~\cite{c2-32} found that although most highly cited scholars working in European institutions had their institutional web pages, they rarely maintained them. Most of them used other social media, which also accelerated the development of Altmetrics.

In addition, reference-related factors and time-related factors have attracted scholars' attention. Dong et al.~\cite{dong2015will} researched scholar's impact considering two reference-related factors: the ratio of max-h-index citations of references to the total number of references of the paper and the average number of citations accumulated by references of the paper. Zhang et al.~\cite{zhang2017exploring} considered academic innovations and assessed scholar impact by a Time-aware ranking algorithm, allocating more credits to the newly published papers according to the representative time functions. Based on the above factors, many evaluation indices have been proposed to quantify scholar impact. In the following two subsections we introduce the counting-based evaluation methods and indices, and network-based evaluation methods and indices respectively.
\subsection{Counting-Based Evaluation Methods and Indices}
\label{sec:2.2.2}
In 2005, Hirsch~\cite{c2-33} proposed the famous h-index to evaluate scholar impact, which is the most famous metric widely used in the whole scientific community. A scholar's h-index means that he has at least $h$ papers cited at least $h$ times. The advantages of h-index include that it is easy to compute and the definition combines quantity and quality of a scholar's outputs. But there are still some scholars who argue that h-index has many shortcomings such as the unbalance between different disciplines, the allocation of co-authors' impact, and the impact of highly cited papers ignored. To keep the impact of highly cited papers from being ignored, Egghe et al.~\cite{c2-34} proposed the g-index. If the citations of all papers published by an author are listed in descending order, the g-index is top $g$ scholarly papers with $g^{2}$ citations.
Similar to the g-index, Jin et al.~\cite{c2-37} proposed R-index and AR-index to overcome the shortcomings of h-index. The R-index is defined as
\begin{equation}
\begin{split}
    R-index = \sqrt{\sum_{i = 1}^h cit_i},
\end{split}
\end{equation}
where $h$ is the author's h-index and $cit_i$ indicates the author's papers that have been cited more than $h$ times, also known as the h-core papers. The AR-index takes age of publications into account, which is calculated by
\begin{equation}
\begin{split}
    AR-index = \sqrt{\sum_{i = 1}^h \frac{cit_i}{a_i}},
\end{split}
\end{equation}
where $a_i$ denotes the $i$-th paper's age.

For the same purpose, Zhang~\cite{c2-35} divided the author's citation function into three parts: the h-squared representing the information of the h-index itself, the excess representing the information of papers having more citations than h-index and the h-tail representing the information of papers with fewer citations. Then, a triangle mapping technique was used to map these three parts to a regular triangle to make the analysis easier. An author's impact was mapped to three parts correspondingly the excess (e-index) representing the research quality, the h-tail (t-index) representing the research quantity and the h-square (h-index) representing the average. This method used three independent parts to quantify an author's impact. In this paper, the authors are divided into two types. The first type of authors have published several high-quality papers but these authors have lower H-index or higher e-index; the second type of authors have published a large number of low-quality papers, but these authors have relatively high h-index, t-index, and lower e-index. Dorogovtsev et al.~\cite{c2-36} developed the o-index to improve the impact of most cited papers. An author's o-index is defined as $o = \sqrt{hm}$, where $h$ is the author's h-index, and $m$ indicates the citations of his/her most cited paper(s).

Another disadvantage of h-index is that it considers all authors of a paper equally. Authors of a multi-authored paper always don't have equal contribution to the work, therefore, the h-index leads to bias. Many studies have tried to solve this problem. Wang et al.~\cite{c2-38} presented A-index to quantify the relative contributions of co-authors. Based on A-index, Stallings et al.~\cite{c2-39}developed a collaboration index, C-index, to quantify the author impact. C-index was defined by
\begin{equation}
\begin{split}
    C-index = \sum_{k = 1}^K A_k,
\end{split}
\end{equation}
where $A_k$ was the author's A-index. The P-index was proposed to quantify researcher's impact by considering the quality of publications, which was given by
\begin{equation}
\begin{split}
    P-index = \sum_{k = 1}^K A_k JIF_k,
\end{split}
\end{equation}
where the $JIF_k$ was the impact factor of the journal where the $k$th paper was published. Besides, some researchers pointed out that even authors that had different citation patterns may get the same h-index. Farooq et al. \cite{c2-40} proposed the DS-index, which is an extension of g-index and intend to provide a distinctive ranking for authors with similar citation pattern. The DS-index is defined as
\begin{equation}
\begin{split}
    DS-index = \sum_{k = 1}^g cit_k,
\end{split}
\end{equation}
where $g$ is the number of g-core papers and $cit_k$ is the $k$th g-core paper's citation. Same as h-core papers, the g-core papers are papers that are used to calculate the g-index of the author.

The indices introduced above are all extension and improvement of h-index. Using h-index can partly reflect the publication behavior and the citation distribution of an author. To more reasonably quantify scholar impact, Sinatra et al.~\cite{c2-42} explored the citation distribution of physicists and found that the highest-impact of a scholar was randomly distributed in their academic careers. Based on this random-impact rule they proposed a stochastic model, in which a unique parameter $Q$ was assigned to predict scholar impact. The Q-value of an author $i$ is calculated by
\begin{equation}
Q_{i}=e^{ \left \langle  \log{c_{i\alpha}} \right \rangle - \mu_{p}}
\end{equation}
where $Q_{i}$ is the $Q$ value of an author $i$. $\left \langle  log c_{i\alpha}\right \rangle$ indicates the average logarithmic citations of all papers published by author $i$. $\alpha$ is the $\alpha$-th paper of author $i$. $\mu_{p}$ is the average impact of luck in the success of papers.

Citation-based author impact evaluation methods show differences among disciplines. Waltman et al.~\cite{c2-47} found that using the fractional counting method can give a more suitable result for cross-field scholar evaluation. Radicchi et al.~\cite{c2-44} proposed a universal variant h-index to solve this problem, named $h_f$-index. In 2013, together with Radicchi, Kaur et al.~\cite{c2-45} improved the $h_f$-index and proposed a new method to compare scientific impact across disciplinary boundaries. The new $h_s$-index was introduced in their work, which was a normalized h-index by the average h-index of all authors in the same disciplines. Lima et al.~\cite{c2-46} considered that a paper can belong to several research areas and the author's impact in an area was calculated by the papers published in the area, which was used by the author's percentile rank. Finally, the impact of an author was quantified by summing up impact across all areas. By this method, although the bias among different disciplines can be reduced, the authors who are active in a rapidly developing area can also get a higher score than others in the basic disciplines.
\subsection{Network-Based Evaluation Methods and Indices}
\label{sec:2.2.3}
Because counting-based evaluation methods are easily manipulated in evaluating scholar impact, scholars explore the structured methods to overcome the shortcomings. The network-based evaluation methods of scholars have evolved from homogeneous scholarly networks to heterogeneous scholarly networks \cite{c2-19,c2-20,c2-21,c2-22,c2-23,c2-24,c2-25,c2-26}. The scholarly networks are made up of academic entities, including scholars, papers, journals or conferences, and institutions. Ding et al.~\cite{ding2009pagerank} used the PageRank algorithm to quantify the impact of the author based on author co-citation network. Yan et al.~\cite{c2-53} developed P-Rank, which used three different networks, including citation network, authorship network, and publish-relationship network, to evaluate the impact of authors, papers and journals. A HITS-type method was first performed to update the scores of papers, authors, and journals in the authorship network and publish-relationship network. Then these scores were used as nodes' initial values to run a PageRank in the citation network to get the final scores of papers. Because the HITS-type algorithm is more suitable for heterogeneous academic networks, mining the academic relationships of heterogeneous networks in depth can make the HITS-type algorithm work better.
Amjad et al.~\cite{c2-55} considered the topic distributions of scholarly entities that were generated by Latent Dirichlet Allocation (LDA) \cite{c2-54} and proposed a topic-based ranking method called Topic-based Heterogeneous Rank (TH Rank). Because of the network complexity and the cost of computing LDA, TH Rank is not an efficient algorithm. Li et al.~\cite{c2-56} put forward a method named QRank for the purpose to rank authors effectively and efficiently. Nykl et al.~\cite{c2-57} used the PageRank algorithm together with several individual evaluation indices, including h-index, publication count, citation count, and author count of a publication to rank scholars.

Although the existing network-based evaluation methods have achieved certain results, the existing evaluation methods still have the following problems: (1) most previous studies quantify author impact based on the first-order academic networks; (2) the citation inflation influences the real impact of the author; (3) the origin of the academic success genes is unknown. Therefore, the higher-order academic network analysis, author impact inflation, and academic success gene need to be explored.
\section{Evaluation of Journal Impact}
\label{sec:2.3}
The impact of journal generates from papers published. Authors are more willing to publish papers on the journal with a high impact. The evaluation of journals is associated with the evaluation of papers and authors. There are several famous publishing groups around the world. They are Elsevier, Springer, Wiley, Wolters Kluwe, and Pearson. It is worth mentioning that the most famous journals, Lancet and Cell, are published by the Elsevier, and Nature is published by the Macmillan. From 1975, Journal Citation Reports (JCR) started to provide the last year's Impact Factor (IF) of journals, together with other evaluation indicators such as the journals' current rank, abbreviated journal title, International Standard Serial Number (ISSN), total cites, immediacy index, total article and cited half-life. Since JCR was taken as an important data resource for quantifying journals. The JCR metrics have become the most popular indices to evaluate journals, and several other metrics have been proposed by the Thomson-Reuters, such as EigenFactor Score (EF), the yearly JCR, and CiteScore\footnote{https://www.scopus.com}. Nowadays, many other evaluation methods and metrics have been proposed except the JCR metrics. In the following subsection, we will discuss the evolution of the existent methods and indices, and summarize the issues of these journal evaluation methods. One possible solution is to explore journal impact inflation and the higher-order academic network analysis.
\subsection{Factors Influencing the Impact of Journals}
\label{sec:2.3.1}
Some classical high-impact journals always have been lasting for many years, such as \emph{Nature} and \emph{Science}. Journal's quality is decided by the quality of papers published on it. Many metrics for evaluating journals are based on citations. The development of the Internet has promoted the paper's citation, as well as the impact of journals. Therefore, open access journals may have a higher impact than the private ones.

Journal impact is with strong discipline, that is, different disciplines have different authoritative journals. Besides, the journal's type may influence its impact factor. Some journals prefer to publish review papers, and some others publish long research papers and short papers. Generally, a review journal impact factor is higher than other journals in the same discipline.
\subsection{The Journal Citation Reports}
\label{sec:2.3.2}
The Journal Citation Report started in 1975. Now it provides more than 10,000 high-quality journals rank every year and is released on the Web of Science (WoS). The evaluation of journal impact contains several usually used metrics, such as journal's total cites, journal impact factor, impact factor without journal self-citations, 5-year impact factor, immediacy index, cited half-life, citing half-life, Eigenfactor score, article influence score and number of citable items of the journal and other metrics. This report is always seen as the most authoritative assessment of journals.

Journal impact factor, which always refers to the 2-year impact factor, was proposed by Garfield in 1955 \cite{c2-60}. The JIF of a journal in year $n$ was defined as follows:
\begin{equation}
\begin{split}
    2-JIF_n = \frac{P_{n - 1} + P_{n - 2}}{C_{n - 1} + C_{n - 2}},
\end{split}
\end{equation}
where $P_{n - 1}$ is the number of papers published on this journal in year $n-1$ and $C_{n - 1}$ is the number of the journal's citations in year $n-1$. The computation of the 5-year journal impact factor is the same as the 2-year impact factor, which is considering the number of papers and citations of the journal in recent 5 years. The impact factor without journal self-citations eliminates the influence of the journal's self-citations, which gives a more objective evaluation of the journal's impact. The cited half-life is years that are taken to reach half of the total citations of the journal, which indicates the persistence of a journal's impact. The citing half-life is defined as the years for the number of references accumulating to half of all, which indicates the novelty of the references.

Other metrics, such as immediacy index, Eigenfactor score, and article influence score, are to cover shortages of the impact factor. The immediacy index is defined as the average citation of papers published on journals in the given year, which can reflect the impact of the journal in that year. Eigenfactor score is calculated by the journal's citation network without self-citation, using a PageRank-type method \cite{c2-61}.
\subsection{Analysis and Improvement of the JCR}
\label{sec:2.3.3}
Although the JCR metrics are used widely, it leads to bias if only using a single metric to assess journals. Many efforts have been paid to overcome the shortages and many other metrics have been proposed, such as H-index for journals \cite{c2-62}, SCImago Journal Rank (SJR) \cite{c2-63}, Source Normalized Impact per Paper (SNIP) \cite{c2-64}. In addition to using a single metric, it is found that the ranking result can be improved by combining these common metrics in some ways, like computing their harmonic means \cite{c2-65} or using the Neural Network to find a non-linear represent \cite{c2-66}. Serenko et al.~\cite{c2-67} found that scholars always preferred to the familiar journals and gave them a higher evaluation. It suggests that introducing personal opinions in the evaluation of journals may be helpful. Tsai et al.~\cite{c2-68} studied the correlation between subjective evaluation (scholars' personal opinions) and objective evaluation (journal rank by JIF and h-index) and used the Borda counting method to combine the two ranking results. Beets et al.~\cite{c2-69} ranked accounting journals referencing the departmental journal lists, which were used to evaluate faculty publications in several famous business schools.

There are also many scholars concern about the relationship among the different journal rank by these metrics \cite{c2-70,c2-71,c2-72,c2-73,c2-74,c2-75}. Setti~\cite{c2-73} argued that it was impossible to capture the real impact of journals by any single indicator. Different evaluation methods quantify journals from different views, so which metrics are more useful is always based on application scenarios. Sometimes it is meaningful to rank journals only by the percentage of highly cited publications of a journal \cite{c2-76}. Besides, the evaluation of journals in different disciplines or different fields of the same subject also needs discussion \cite{c2-76,c2-77}.

Chatterjee et al.~\cite{c2-78} studied the citation distribution and found that a few high-cited papers had hold most citations in both journals and institutions. Based on many research of the citation distribution of journals, Kao et al.~\cite{c2-79} proposed a stochastic dominance analysis based method to evaluate journals.
\subsection{Network-Based Evaluation Methods and Indices}
\label{sec:2.3.4}
The most used methods for evaluating nodes in network are PageRank and HITS. As discussed in the previous sections, HITS algorithm can be used to rank paper, author and journal together. There are some PageRank-type methods being designed for ranking journals, which have a basic form like
\begin{equation}
\begin{split}
    r(J_i) = (1 - \lambda)x_i + \lambda \sum_{j} [r(J_j) \times \frac{w(J_j,J_i)}{sum_{k} w(J_j,J_k)}],
\end{split}
\end{equation}
where $x_i$ indicates the adaptive damping factor and satisfies $\sum_{i = 1}^{N} x_i = 1$. Generally, the value of $x_i$ is set as $\frac{1}{N}$. $r(J_i)$ represents the importance score of journal $i$ \cite{c2-80}.

Based on the PageRank algorithm, Chen~\cite{c2-81} added the expert judgments on the method as a weight part, and optimized the function by Particle Swarm Optimization (PSO). In the same way, Lim et al.~\cite{c2-82} used the relevance and importance of the citations between journals to design the weighted PageRank. Zhang proposed the HR-PageRank algorithm to evaluate journal impact via weighted PageRank according to the author's H-index, and relevance between citing and cited papers \cite{Zhang2017Evaluating}. Bohlin et al.~\cite{c2-83} studied the different performances of zero- (the classical Markov model), first- and second-order Markov model while ranking journals and found that higher-order Markov models performed better and were more robust.

Some evaluation methods consider the structural position of journals in the journal citation network. Zhang et al.~\cite{c2-84} proposed an indicator named Quality-Structure Index (QSI), which ranked journals by the intrinsic popularity and structural position of journals. The intrinsic popularity was quantified by some frequently used metrics, such as JIF, Eigenfactor score, PageRank score. Similarly, Leydesdorff~\cite{c2-85} introduced the betweenness centrality of journals in the journal citation network to the assessment task. Su~\cite{c2-86} gave a link-based representation to some frequently used metrics for journals, such as JIF, and proposed a link-based fusing method to fuse several metrics together according to the links in and among paper citation network, authorship network and paper publishing network. This method has found a new way to consider many metrics together to evaluate academic entities.

Based on the above analysis, the existing journal evaluation methods still have the following problems: (1) most previous studies quantify journal impact based on the first-order academic networks; (2) the citation inflation influences the real impact of a journal. Therefore, researchers need to explore the higher-order academic network analysis and journal impact inflation to resolve the challenging issues of journal evaluation.
\section{Open Issues and Challenges}
\label{sec:4}
In this section, several open issues and challenges are shown for further research in this area, including the pattern of collaboration impact, unified evaluation standards, implicit success factor mining, dynamic academic network embedding, and scholarly impact inflation.
\subsection{Pattern of collaboration Impact}
A significant amount of work has been focused on quantifying the impact of scholarly papers, scholars, and journals~\cite{bai2018quantifying,c2-42,Zhang2017Evaluating}. However, little is known about how the impact of scientific collaboration evolves over time. Previous researchers measure the impact of co-authors by citations, which are easy to be manipulated. The structured methods for measuring the impact of co-authors are urgently needed in the science community. With the available large-scale datasets on citations and collaborations, it becomes possible to explore the patterns of collaboration impact in scientific collaborative careers over time and their potential relationships with scientists' success. Since the structured methods are needed to quantify the impact of co-authors, how to construct the network to measure the collaborative impact and how to model remain the broader challenge. One possible solution is to construct a heterogeneous academic network in which the impact of the co-authors are quantified. Based on this, researchers explore the pattern of collaboration impact.
\subsection{Unified Evaluation Standards}
We have mentioned many automatic evaluation methods that try to find high-quality papers from a mass of publications. But these methods can only give researchers suggestions that which paper may be useful, the contents of the papers recommended are not concerned by the algorithm. Therefore, there is still a strong demand for efforts to find the papers you need in the research process. Although there are many automatic evaluation methods, we can not find a unified evaluation standard to evaluate which method can outperform others. A widely accepted ground truth is of great need in the evaluation systems. To solve this problem, the data set must be unified first.
\subsection{Implicit Success Factor Mining}
In the past, more attention has been given to explicit success factors. In author impact evaluation research, researchers have found some explicit success factors such as academic age, institution, research field, and country~\cite{cai2019scholarly}. However, little is known about the mechanisms of the temporal evolution of success in science. Uncovering the origin of the success factors in science is a challenging task. Success in science may depend on exogenous factors, such as mentor-student relationship, learning habits, and education level, remains unknown. Actively exploring the relationship between exogenous factors and academic success may provide a method for implicit success factor mining.
\subsection{Dynamic Academic Network Embedding}
Many static network embedding methods have been proposed, however, academic networks evolve over time. For example, in citation networks, citing papers and cited papers always dynamically change over time, e.g., new citations are continuously added to the citation networks when authors cite previous research work. To learn the representations of nodes in dynamic scholarly networks, the existing academic network embedding methods need to run repeatedly and take time. Therefore, further study on dynamic scholarly network embedding algorithms remains an open challenge in this area. To obtain the efficient representation, a deep feature learning and the associated representation model supported by dynamic academic data may need to be established.
\subsection{Scholarly Impact Inflation}
Scholarly impact inflation, which arises from the exponential growth of scholarly papers, affects the real value of scholarly impact, therefore, impacting the comparative evaluation of papers, scholars, journals, institutions, and country output across different periods~\cite{pan2018memory}. Scholars can increase their citations by relying on their friends and co-authors, indicating that citations are easily manipulated. Many work  has focused on unraveling the dynamics of inflation for citations~\cite{higham2017unraveling,petersen2019methods,Bai2016Identifying,tarkhan-mouravi2020traditional}. Under the background of the inflation of citations, how to construct the evaluation network of scholarly impact and how to model are surprisingly difficult, highlighting the broader challenge of evaluating the scholarly impact in the science community. One possible solution is to weaken citation inflation through the higher-order academic networks.
\section{Conclusion}
\label{sec:5}
In this paper, we conduct a comprehensive review of the literature in quantifying success in science, focusing on evaluation indices of scholarly impact. Two changes have taken place in quantifying success in science research: (1) from unstructured evaluation indices to structured evaluation indices; (2) from single-disciplinary impact assessment to interdisciplinary impact assessment. However, the literature-based analysis has led to the conclusion that despite a large number of evaluation indices have been used to resolve the problems in quantifying success in science, the solutions of some potential issues remain unknowns, such as the pattern of collaborative impact, implicit success factor mining, dynamic academic network embedding, and scholarly impact inflation. To solve these challenging issues, researchers can explore from the high-order scholarly network, heterogeneous network analysis and modeling, and academic relationship identifying.

\bibliographystyle{IEEEtran}

\EOD

\end{document}